\documentclass[sn-chicago,referee]{sn-jnl}

\usepackage{graphicx}%
\usepackage{multirow}%
\usepackage{amsmath,amssymb,amsfonts}%
\usepackage{amsthm}%
\usepackage{mathrsfs}%
\usepackage[title]{appendix}%
\usepackage{xcolor}%
\usepackage{textcomp}%
\usepackage{manyfoot}%
\usepackage{booktabs}%
\usepackage{algorithm}%
\usepackage{algorithmicx}%
\usepackage{algpseudocode}%
\usepackage{listings}%

\usepackage{tabularx}
\usepackage[table,dvipsnames]{xcolor}
\arrayrulecolor{white}
\usepackage[T1]{fontenc} 

\begin{document}

\title
[Telepresence Robot Deployments with Homebound \mbox{K-12}
Students]
{Multi-Week, In-Class Deployments of Telepresence Robots With Four Homebound \mbox{K-12} Students: Benefits, Challenges, and Recommendations}


\author*[1]{\fnm{Matthew} \sur{Rueben}}\email{rueben@up.edu}

\author[2]{\fnm{Rhianna} \sur{Lee}}\nomail

\author[1]{\fnm{Thomas R.} \sur{Groechel}}\email{groechel@usc.edu}

\author[1]{\fnm{Hengzhi} \sur{Chen}}\email{realchen@usc.edu}

\author[3]{\fnm{Haemi} \sur{Lee}}\email{hlee22@colby.edu}

\author[4]{\fnm{Gisele} \sur{Ragusa}}\email{ragusa@usc.edu}

\author[1]{\fnm{Maja J.} \sur{Matarić}}\email{mataric@usc.edu}

\affil*[1]{\orgdiv{Department
of Computer Science}, \orgname{University of Southern California}, \orgaddress{\city{Los Angeles}, \state{CA}, \country{USA}}}

\affil[2]{\orgaddress{\city{Los Angeles}, \state{CA}, \country{USA}}}

\affil[3]{\orgname{Colby College}, \orgaddress{\city{Waterville}, \state{ME}, \country{USA}}}

\affil[4]{\orgdiv{Division of Engineering Education, Viterbi School of Engineering}, \orgname{University of Southern California}, \orgaddress{\city{Los Angeles}, \state{CA}, \country{USA}}}




\abstract{
Missing significant amounts of school during K-12 education is known to put students' cognitive and social development at risk. 
Alternatives such as home instruction and online learning are common, but lack sufficient interaction with peers and teachers in the classroom.
Mobile remote presence systems, or telepresence robots, are promising for homebound students because they provide embodiment and mobility in addition to the real-time participation offered by video conferencing technologies. 
Research is needed, however, for telepresence robots to meet the complex needs of homebound students participating remotely in the \mbox{K-12} classroom context.
We present findings from four multi-week deployments with homebound \mbox{K-12} students attending classes via  telepresence robots. 
The homebound students' experiences were documented in a total of 15 interviews and analyzed qualitatively as case studies.
The homebound student participants and their deployment contexts differed from one another along multiple dimensions, and while some benefits of mobile remote attendance were enjoyed by all participants, each participant also experienced unique benefits.
Some challenges with hearing, seeing, and moving the robot around the classroom warranted improvements to the design of the telepresence system. 
Other challenges suggested priorities for managing a classroom deployment, such as ensuring that the remote student is included in classroom activities, accountable to the teacher, and treated with respect by classmates.
Based on insights from the study, we make recommendations for real-world deployment procedures in similar contexts. 
}




\maketitle

\section{Introduction}
In 2019, at least 10\% of 6--17 year old children in the United States missed seven or more days of school \citep{usdhhs_2019}.
Various reasons can prevent students from being able to physically attend school, such as chronic health conditions \citep{charlton1991absence,hill1989asthma}, short-term illness \citep{neuzil2002illness}, and anxiety and behavioral disorders \citep{egger2003school,mcshane2001characteristics}. 
Solutions such as home instruction, individualized tutoring, and online learning are common, but include little or no interaction with peers and teachers in the classroom, both essential components of cognitive and social development for \mbox{K-12} students \citep{national2000people,mayer2003learning}. 
Video conferencing technologies have made it feasible for remote students to have live, audiovisual interactions with people in the classroom.
Remote attendance presents many challenges, however, as was seen during the COVID-19 pandemic \citep{huck2021effects}.

This work focuses on the promise of mobile remote presence systems, or telepresence robots, which can endow remote students with a physical embodiment in the classroom, allowing them to move around, thereby also controlling the origin and direction of their gaze \citep{kristoffersson_review_2013,fels_telepresence_2001}.
We refer to this experience as {\it mobile remote attendance}. 
Research on telepresence robots for \mbox{K-12} mobile remote attendance began around 2001 \citep{fels_telepresence_2001,weiss_pebbles_2001}, and has grown significantly more recently \citep{hafner2022limits,page2021telepresence}. 
Telepresence robots were designed for workplace settings \citep{kristoffersson_review_2013,tsui2013design} that largely lack some of the ubiquitous challenges of \mbox{K-12} classrooms, notably noise, clutter, and bullying \citep{newhart_virtual_2016}. 
In the case of home- or hospital-bound students, the home or hospital environment is also involved due to the two-way nature of the audiovisual connection \citep{newhart2017my}.
Students have a wide variety of medical contexts, accessibility needs, personalities, preferences, and past experiences with in-person education \citep{rueben2021long} that impact mobile remote attendance and make this a challenging research problem. 

This work was part of a larger research project studying how telepresence robots can improve the outcomes of homebound \mbox{K-12} students during extended absences from school \citep{rueben2021long,cha_designing_2017,fitter_are_2020,cha2016enabling,cha_my_2017,fitter_evaluating_2018,fitter2020closeness}.
Here we present findings from four multi-week deployments, each with a different homebound \mbox{K-12} student at a different school. 
Two of the participants were homebound due to medical conditions that also presented accessibility challenges for using remote presence robots; the other two participants were homebound due to mental health concerns. 
Data collection was focused on the homebound students' experiences; we conducted a total of 15 interviews along with audiovisual recordings.  
We performed a qualitative analysis of each deployment as a case study, followed by a cross-case analysis. 
Our results contribute: profiles of four homebound \mbox{K-12} students, their accessibility needs, and the four deployment contexts; the benefits and challenges they experienced while using the robots; recommendations for telepresence robots and their interface design; and recommendations for adequately managing such deployments.

\section{Motivation}
Prolonged absence from school can negatively affect academic performance \citep{gottfried2011detrimental} and lead to increased dropout rates \citep{bridgeland2006silent}.
Classroom activities not only promote learning but also aid in the development of important social skills \citep{johnson1986cooperative,johnson2000cooperative}. 
It is therefore important to minimize the effects of physical separation from school and allow students to engage in regular classroom activities. 
The need for effective remote learning became critical during the COVID-19 pandemic, motivating global initiatives by organizations such as UNESCO \citep{unesco_2020}.

Typical remote learning solutions such as tutoring, home instruction, and online learning lack peer-mediated educational experiences \citep{bransford2006learning}---students usually cannot actively engage and connect with their peers, participate in group work, and contribute to class discussions, among other common classroom activities. 
Technologies such as video conferencing offer remote students a view of the classroom from a fixed perspective, limiting opportunities for participation; remote students do not always feel present or able to express themselves in the classroom \citep{fitter_are_2020}.

\subsection{Telepresence Robots are Promising for Mobile Remote \mbox{K-12} Classroom Attendance}
A mobile remote presence system \citep{lee__2011} allows the user to perform two-way video conferencing and navigation in a remote environment \citep{kristoffersson_review_2013}. 
The two main benefits of mobile remote presence platforms are (1) enhanced user independence and (2) a greater feeling of presence, as the user is able to move about the remote environment and interact with people there \citep{kristoffersson_review_2013}.

As early as 2001, in-school deployments showed the promise of mobile remote attendance in helping remote students to keep up with course material and to contribute to classroom discourse \citep{fels_telepresence_2001,weiss_pebbles_2001}. 
The foundational PEBBLES system allowed for two-way audio and video conferencing between a hospital-bound student and their \mbox{K-12} classroom \citep{weiss_pebbles_2001}.
Results from five-, six-, and nine-week deployments revealed that, ``remote students were able to engage in the same tasks as their peers (with minimal disturbances to their concentration), to initiate positive contributions to the classroom, and to communicate'' \citep{fels_telepresence_2001}.
However, few telepresence robot systems have been developed specifically for attending school \citep{kristoffersson_review_2013}.

\subsection{Research in the \mbox{K-12} Classroom Context is Needed}
Telepresence robots require assessment and improvements before they are ready to adequately represent remote students \citep{cha2016enabling}.
Currently, most telepresence platforms are designed for adult users to facilitate communication and collaboration in workplace settings \citep{kristoffersson_review_2013,tsui2013design}. 
The classroom environment and young users present unique design challenges: the \mbox{K-12} classroom environments are diverse and highly dynamic, and student users vary broadly in capabilities and needs \citep{tsui_designing_2011}. 
Furthermore, remote students may attract attention and even bullying while using the robot \citep{fels_telepresence_2001,newhart_virtual_2016}.
The consequences of failure are also severe: disruptions in the learning environment can negatively impact educational outcomes and hinder technology adoption \citep{hu2003examining}.
Even if telepresence robots are a good way for homebound \mbox{K-12} students to remotely attend school, there is a risk that teachers, school administrators, or students' families will not adopt them \citep{newhart2017my}. 
Consequently, it is necessary to consider the unique research and design challenges of applying telepresence robots to the \mbox{K-12} classroom context \citep{cheetham2000interface}.

\section{Related Work}
Recent years have seen a flurry of studies of \mbox{K-12} students attending class via telepresence robots \citep{hafner2022limits}.
Some of the work has focused on the hospital context \citep[e.g.,][]{soares2017mobile}, while other work has addressed homebound students and the classroom context \citep[e.g.,][]{newhart2017my}.
Recent research has also examined particular contexts that keep students out of school, such as chronic illness \citep{page2021telepresence,newhart_virtual_2016} or cancer \citep{weibel2020back,powell2021keeping}. 
Areas of special focus have included technology adoption \citep{newhart2017my}, relationships between stakeholders \citep{johannessen2022student}, and robot and user interface design (see next subsection). 
Several recent studies have also produced themes for future work, such as setting up the environment \citep{newhart2017my}, training teachers \citep{newhart2017my}, expectations \citep{weibel2020back}, and ``one size does not fit all'' \citep{powell2021keeping}.
Some work has taken theory-inspired approaches, such as drawing on science and technology studies (STS) to produce a critical component structure from interview responses \citep{johannessen2022student}.  
Our study complements and extends past work by focusing primarily on the experiences of homebound students and the usability of a telepresence robot and its user interface.
Our participants included students who were homebound due to illness, physical disability, mental health concerns, or a combination of these factors. 

\subsection{Design of the Telepresence Robot and User Interface for \mbox{K-12} Education}
Some previous work has focused on the design of telepresence robots and user interfaces for mobile remote attendance in \mbox{K-12} education. 
\cite{cha_my_2017} presented an interactive game that provided a virtual classroom environment to \mbox{K-12} students for prototyping and initial evaluations of system designs.
 \cite{cha_designing_2017} also conducted field sessions in which four designers operated a telepresence robot in a \mbox{K-12} classroom to collect design insights. 
\cite{ahumada2019going} gathered insights on user needs through a study of 19 homebound students attending school via a telepresence robot.
\cite{fitter_evaluating_2018} and \cite{fitter2020closeness} studied the effects of personalizing the appearance of remote presence robots. 
\cite{cheetham2000interface} described the design process for a device on PEBBLES \citep{fels_telepresence_2001} for getting attention. 
In another study of the same four deployments described in this article, \cite{rueben2021long} evaluated a feedback system for improving speech intelligibility of homebound students while using remote presence robots. 
In this work, we asked homebound students about the usability of the system, especially with respect to participating in classroom activities and communicating with people in the classroom.
We also gathered design recommendations directly from the homebound students and their families and teachers. 

\subsection{Effects of Mobility and Embodiment in Education}
Past research has investigated the effects of embodiment and mobility afforded by telepresence robots via a direct comparison to video conferencing and other distance learning tools. 
To our knowledge, only one such study has been published with participants at the \mbox{K-12} level \citep{shin2017qualitative}. 

In contrast, several studies have directly compared telepresence robots to other distance learning technologies at the university level. 
One study conducted focus groups with twelve online students who compared using video conferencing, Kubi (tabletop) telepresence robots, and Double (mobile) telepresence robots \citep{gleason_hybrid_2017}. 
Another team performed a similar study with iPads on stands as an additional option \citep{bell_2d_2016}.
A study of students in four university courses compared preferences and attitudes of attending in person, via state-of-the-art distance learning tools, and via a telepresence robot \citep{fitter_are_2020}. 

The work presented in this paper did not directly compare telepresence robots with video conferencing or other technologies that could be used for remote attendance, as that has been covered in past work described above. However, some of the interview prompts were designed to explore the effects of the robot's embodiment and mobility. 

\section{Approach}
\label{sec:approach}
This work is part of a larger project that studied how telepresence robots can improve the outcomes of homebound \mbox{K-12} students during extended absences from school \citep{rueben2021long,cha_designing_2017,fitter_are_2020,cha2016enabling,cha_my_2017,fitter_evaluating_2018,fitter2020closeness}.
This paper describes a study focused on homebound students' experiences (1) using the system, (2) being telepresent in the classroom, and (3) interacting remotely with classmates and teachers. 

Our study design was guided by several concepts of interest for each of these three types of experiences.
For experiences using the system, we focused on user needs, usability in a range of classroom use cases, and recommendations for new features. 
For experiences from being in the classroom, we focused on feelings of presence and self-consciousness, engagement, and the ability to participate in classroom activities.
For experiences interacting with classmates and teachers, we focused on social inclusion, the ability to get or avoid attention, conversations, and relationships. 

In spite of significant logistical complexity, we conducted multi-week deployments to capture (1) the initial setup and deployment, (2) the first few class sessions of use, and (3) several more weeks of participants, classmates, and teachers becoming acclimated to the system.

Both self-report and behavioral measures were used. 
Interviews were used for gathering in-depth reports from participants, focusing on the concepts of interest described above.
Audiovisual recordings were used to verify participants' stories, add details, and confirm the order of events. 

Data analysis was qualitative to identify themes, and was performed by study team members who knew the people and context of each deployment. 
Because each participant and deployment context were quite different from the others in several ways (see Table~\ref{tab:overview}) and we had a sample size of only \(n=4\) participants, we analyzed each participant's deployment as an independent case instead of attempting to make inferences about the population. 
These case studies were followed by a cross-case analysis to compare results between participants. 

\section{Methods}
\label{sec:methods}
We conducted four deployments of a telepresence robot system, each with one homebound \mbox{K-12} student participant.
For privacy, in the rest of the paper we refer to the participants by using they/them pronouns and the identifiers P1, P2, P3, and P4 according to the chronological order of the deployments.
This section presents the details of the telepresence robot system, deployment procedure, measurement methods, and data analysis processes.  
This study was approved by the Institutional Review Board (IRB), study number UP-18-00129.

\begin{figure}[!t]
\centering
\includegraphics[width=11em]{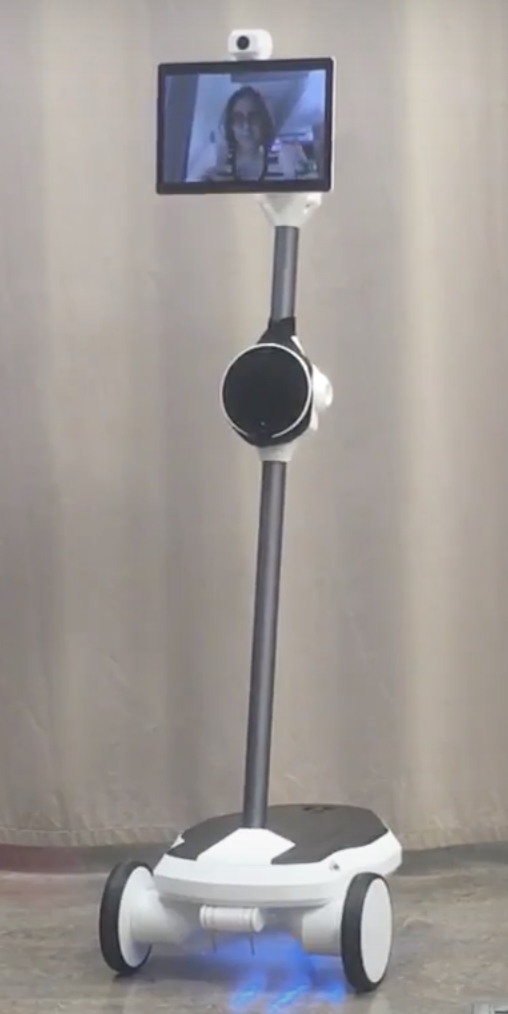}
\caption[OhmniLabs telepresence robot used in the deployments, shown in the research lab, being operated by a lab member]{OhmniLabs telepresence robot\protect\footnotemark[1] used in the deployments, shown in the research lab, being operated by a lab member}
\label{fig:ohmni}
\end{figure}

\subsection{The Mobile Remote Attendance System}
Participants attended class using OhmniLabs telepresence robots,\footnotemark[1] as shown in Figure~\ref{fig:ohmni}.
The Ohmni robot uses a 10.1-inch touchscreen to display a video of a remote operator's face. 
A webcam is mounted above the display. 
A speaker-and-microphone unit is mounted below the display, equipped with buttons for adjusting the volume and muting the remote operator's voice. 
A wheeled base allows the remote operator to drive the robot forward, backward, and to rotate in place; the remote operator can also tilt the webcam (along with the display) up or down. 
We attached a foam border to the display to protect it in case the robot was bumped or knocked over.
We also supplied classroom teachers with a bicycle lock for securing the robot. 

\footnotetext[1]{The OhmniLabs website: \url{https://ohmnilabs.com/products/ohmni-telepresence-robot/}.  The website shows a newer version of the robot than was used in this work.}

Each homebound student was loaned a MacBook Air laptop to access the robot's web-based operator interface. Each participant either used in-ear headphones with a built-in microphone or the laptop's built-in microphone and speakers. 
The video feed of the student operator's face was captured by the laptop's built-in webcam. 
We used the OhmniLabs web-based user interface, which included a slider for adjusting the volume of the operator's voice from the robot's speakers. To turn the robot, the operator could click anywhere on their view of the robot's webcam feed and the robot would turn to recenter its view on that point. The interface included buttons for muting the operator's audio and turning off the video feed of their face to the robot's screen. 


P1 asked that their face not be shown on the robot's screen, so we implemented a button for displaying an image of their choice instead. 
P4 also used this feature; the details for both deployments are discussed in Section~\ref{sec:seeing-and-being-seen}. 


Further accommodations were made to the user interface to make it accessible for P1 and P3, with help from teachers on special assignment (TOSAs) from the school district who specialized in educational technology. 
Accommodation details can be found in P1's and P3's deployment overviews, in Section~\ref{sec:deployment-overviews}. 

\subsection{Deployment Procedure}
\label{sec:deployment-procedure}
All participants were recruited from a large urban school district in the US. 
Each deployment occurred in academic years 2018-19 and 2019-20 and lasted multiple weeks (see Table~\ref{tab:overview}).

\subsubsection{Initial Setup, Training, and Classroom Introduction}
\label{sec:setup}
Members of the study team visited each homebound student and family at their home to perform setup and training for the robot system. 
First, the participant's home wireless network was configured to work with the robot control laptop via a direct Ethernet connection to the router where possible.
The participant and, as applicable, family members who would be helping the participant use the robot, were then trained on the robot's user interface and practiced driving the robot within the same room.

Members of the study team also visited the school prior to the start of the study deployment to demonstrate the robot to teachers and, in some cases, other school administrators. 
We showed teachers how to charge the robot, troubleshoot basic issues (e.g., by restarting the robot), and contact the study team for help. 
We worked with teachers to determine a suitable location for the robot's charging station, which was also where the robot was stored (and locked) when not in use. 
We worked with school district information technology staff to register the robot(s) being deployed with the school district's wireless network; each robot was given access to a network with bandwidth priority over the one used by other students.

After the setup and training, the robot was placed in the classroom and the teacher(s) worked with the participant and family to choose the day of the first class session the participant would attend remotely.
One or more classes were identified for the participant to remotely attend (see Table~\ref{tab:overview}). 
For the remainder of the study, the participant and their family had the freedom to decide whether to attend each class session as planned.  

\subsubsection{Ongoing Deployment Operations}
The study team also provided technical support throughout the deployments. 
Participants and their family members were encouraged to call or text the lead author directly with any technical difficulties.
Teachers were offered the same service.
Requests for technical help tended to be frequent at the beginning of a given deployment and decreased with time.


The deployment ended with the robot(s) being retrieved from the classroom(s) and the laptop from the participant's home. 
Each of the four homebound student participants was provided a 100 USD Amazon gift card as compensation. 
Each classroom teacher was also provided a 100 USD Amazon gift card, as were P1's caretaker brother and the two TOSAs who helped with recruitment and coordination of the teachers and homebound student participants.
The student peers in each of the classrooms were provided with pencils and lanyards bearing the logo of the study team's institution.



\subsection{Interviews}
\subsubsection{Data Collection}
We conducted interviews with the homebound student participant at least once per week.
Interviews lasted about 30--60 minutes, and were audio-recorded; the interviewer was a research team member with experience working with children as research participants. 
The interviewer went to the participant's home for all interviews except for the last two with P4 (Weeks 5 and 6), which were conducted by phone. 
Family members were sometimes included in the interviews. 
P1's brother and P3's mother and brother were included whenever possible because they were helping the participants use the telepresence system as well as helping them through the interview process. 

Interviews were semi-structured---the interviewer asked questions from a protocol, and had the discretion to ask follow-up questions. 
The protocol for each interview was assembled from a pool of questions created by the study team and organized into seven sections based on the concepts of interest described in Section~\ref{sec:approach}.
Appendix~\ref{append:interview-question-pool} provides the full pool of questions in each of the seven sections.
Questions from all of the following sections were used in each interview: General Impressions, Relationships with Classmates, The Robot's Technical Abilities, and Goals and Regrets.
Additionally, 1--2 of the following sections were chosen before each interview: Interactions with Classmates, Learning Experience, and Ideas for Improving the Robot or Interface. 
For each section chosen for a particular interview, questions from that section were chosen, reviewed, and sometimes modified before the interview. 
Choices about which sections and questions to include, as well as any modifications to questions, were made by the study team in response to what was happening in the classroom and to track phenomena of interest. 

\subsubsection{Data Analysis}
The interview data were subjected to a qualitative analysis comprised of a case study of each participant followed by a cross-case analysis.
First, audio recordings of the interviews were transcribed by the interviewer. 
The interviewer then wrote detailed notes about each interview, especially where a respondent's meaning was unclear from the transcript of the audio recording. 

As described in Section~\ref{sec:setup}, the lead author had visited each home and school and met each participant, their teacher(s), and any family members involved in the deployment.
The lead author also compared some of the transcribed interview responses to the audiovisual recordings to verify the respondent's account and to recover key details or the order of events. 
The lead author then produced the following results from close readings of the interviewer's transcripts and notes: 
\begin{itemize}
    \item An overview of each multi-week deployment (Section~\ref{sec:deployment-overviews}), divided into a profile of the participant, notes on any accessibility needs, and other context.
    \item A list of benefits of and challenges with the remote attendance system, as well as potential solutions to those challenges (Sections~\ref{sec:benefits} and~\ref{sec:challenges-and-solutions}). These were identified for each participant in their individual context (as part of their case study), then compared between participants (as part of the cross-case analysis). Benefits and challenges were selected for reporting when they were stated in the interviews explicitly, clearly, and with their impact either explicitly stated or implied by emphasis or repetition, even if they were unique to a specific participant. The selected benefits and challenges were then grouped by similarity into sections for reporting. The potential solutions were all suggested or implemented by the study team, participants, family members, classmates, teachers, or TOSAs. 
\end{itemize}

\subsection{Audiovisual Recordings}
\subsubsection{Data Collection}
We aimed to collect an audiovisual recording for the entire duration of every mobile remote attendance session. 
Recordings were made from two perspectives: (1) the camera and microphone on the control laptop recorded the homebound participant's face and voice as it would be presented to the classroom by the robot; and (2) the robot's camera and microphone recorded the classroom as it was presented to the homebound participant on their user interface. 
The homebound participant or their caretaker was responsible for starting and ending the two recordings; we partly automated this process via custom scripts and macros. 

\subsubsection{Data Analysis}
For each mobile remote attendance session, the two recordings were synchronized so they could be viewed simultaneously. 
The audiovisual recordings were used when a participant's interview response was ambiguous or did not specify important details such as the exact order of events.

\section{Results}
\label{sec:results}

\subsection{Overview of Data Collected}
Table~\ref{tab:overview} provides information about the deployments. ``Week~1'' indicates the week when a participant first remotely attended class. P3 did not attend class on Week~5 due to a holiday, and P4 did not attend class on Week~3 due to illness. 

\begingroup
\renewcommand*{\thefootnote}{\textit{\alph{footnote}}} 
\begin{table}[ht]
\caption{Information about each participant's deployment, including medical context and data collection. }
\label{tab:overview}
\centering
\begin{tabularx}{\textwidth}{
>{\bfseries\hsize=0.25\hsize}X |
>{\hsize=0.15\hsize}X |
>{\hsize=0.15\hsize}X |
>{\hsize=0.15\hsize}X |
>{\hsize=0.15\hsize}X
}
\rowcolor{BrickRed}
 & \textcolor{white}{\textbf{P1}} & \textcolor{white}{\textbf{P2}} & \textcolor{white}{\textbf{P3}} & \textcolor{white}{\textbf{P4}} \\
\hline
\rowcolor{Dandelion}
Level & High School & High School & High School & Middle School \\
\hline
\rowcolor{Dandelion}
Classrooms & Spanish & History, English, Law & Intensive \mbox{Studies} & Math, Science, English, History, Lunch \\
\hline
\rowcolor{Dandelion}
Duration & 4 weeks & 2 weeks & 8 weeks & 5 weeks \\
\hline
\rowcolor{Dandelion}
Days Used Robot\footnotemark[1] & 10 days & 9 days & 23 days & 14 days \\
\hline
\rowcolor{GreenYellow}
Difficulty Speaking? & Yes & No & Yes & No \\
\hline
\rowcolor{GreenYellow}
Difficulty with Mouse or Keyboard? & Yes & No & Yes & Some \\
\hline
\rowcolor{YellowGreen}
Interview Weeks & 1, 2, 3 & 1, 1, 2, 2 & 2, 4, 6, 7 & 1, 2, 5 | 6 \\
\hline
\rowcolor{SkyBlue}
Audiovisual \mbox{Recordings} & 14 hours & 10 hours & 15 hours & 28 hours \\
\end{tabularx}
\footnotetext[a]{Number of days in which the participant successfully logged into the robot to attend any part of a class.} 
\end{table}
\endgroup

\subsubsection{Interviews}
Table~\ref{tab:overview} shows the week each interview was conducted; on weeks without interviews, the participant was usually not remotely attending class due to holidays or medical issues. P2 was interviewed twice per week due to the brevity of their deployment, and P4's final interview was the week after the deployment ended.



\subsubsection{Audiovisual Recordings}
\label{sec:recordings-results-overview}
Recordings sometimes failed when a participant restarted the robot user interface or when a laptop ran out of disk space; also, participants sometimes forgot to start recording. 
As a result, one or both of the recordings were missing for some of the sessions. 
A total of 67~hours of audiovisual recordings were successfully collected---see Table~\ref{tab:overview} for a breakdown by participant. 

\subsection{Participant Profiles and Deployment Notes}
\label{sec:deployment-overviews}
There was remarkable variance among the four participants in terms of medical contexts, past experiences in the classroom, and personalities. This section introduces each participant, the adaptations made to the telepresence system for their accessibility, and any additional notes about their deployment not mentioned elsewhere. Further details are shown in Table~\ref{tab:overview}. 

\subsubsection{P1}
\underline{Profile:} P1 was a bit of a jokester. They liked wrestling, sports cars, and video games. One of P1's nurses said that they often stayed up until 5am playing video games, and then woke up very late. This made it difficult to adapt to a normal school schedule without falling asleep during class.

P1 had a progressive medical condition that had confined them to bed with a ventilator prior to the deployment. There was a full-time nurse present, and P1's older brother was a paid guardian. P1 had not attended school in person for about 6 years, and instead had a homeschool teacher.

\underline{Accessibility:} By the start of the deployment, P1 could push buttons or keys on a keyboard, but only if someone placed their hand in the right position. Their first finger could only make a sweeping arc on a trackpad. To drive the robot, P1's brother placed P1's hands in an unconventional position on a Playstation 3 controller\footnote[2]{\url{https://en.wikipedia.org/wiki/DualShock\#DualShock_3}} such that their thumb and first finger could move the joystick and press several of the buttons. The Enjoyable app\footnote[3]{\url{https://yukkurigames.com/enjoyable/}} was used on the robot control laptop to remap some buttons to allow P1 to drive the robot forward and turn. 

Speaking while using the robot was difficult for P1 because they needed to pause for air. They often spoke in spurts of one or only a few words. 
P1 also could not form facial expressions beyond a slight smirk. 

\underline{Deployment Notes:} From the beginning of the deployment, P1 decided not to show their face on the robot's screen, choosing instead to display an image of a sports car (see Section~\ref{sec:seeing-and-being-seen}).

\subsubsection{P2}
\underline{Profile:} 
P2 described themself as a person who often raises their hand and likes to participate in class discussions. 
They left school three months before the deployment, reporting feeling overwhelmed by the people and things to do. 
They reported that negative things about them had been said and posted on social media by other students in the past. 
P2 was scheduled to return to class about three months after the deployment. 

\underline{Accessibility:} No adaptations were made to the telepresence system for P2. 

\underline{Deployment Notes:} P2 remotely attended two classrooms for the entire deployment. Two robots were deployed for P2: one in each classroom.

\subsubsection{P3}
\underline{Profile:} P3 was very vocal and gregarious—their mother said they would talk to anybody, and loved attention and connecting with their teacher and classmates. 

P3 had been in and out of school since the 5th grade due to their medical condition (see below). P3 wanted to go to school, but when they did so they often caught a respiratory illness and were hospitalized. Some of P3's medical treatments made them unenthusiastic and tired. 

\underline{Accessibility:} P3's mother was present for all remote class sessions, and one of their brothers, who had been a teaching assistant in P3's class before, was also often there to help with the technology. An external monitor was set up to supplement the screen of the robot control laptop. Four large buttons were placed at elbow height on P3's wheelchair for moving the robot forward, back, left, and right; prior to that, their mother or brother drove the robot. 

P3 had difficulties speaking that made their voice sound slurred, gurgling, and inarticulate. P3's mother often helped people in the classroom to understand P3 by repeating what she understood P3 to be saying. 

\underline{Deployment Notes:} P3 had to log out from the robot on several occasions due to medical emergencies; they also had a minor surgery that created a gap in the deployment. 

\subsubsection{P4}
\underline{Profile:} P4 described themself as naturally shy. They had a passion and a talent for drawing, and had requested to attend a high school for the arts. 

P4's medical condition made movement painful, including brushing their hair and using a computer.
Prior to their deployment, P4 had experienced other students making jokes about their disability. 

\underline{Accessibility:} No adaptations were made to the telepresence system for P4. 

\underline{Deployment Notes:} P4 started by attending two classes via the robot, and then added more as the deployment progressed. A single robot was deployed for P4 to use in all the classes. Like P1, P4 chose not to show their face on the robot's screen, instead displaying one of their drawings (see Section~\ref{sec:seeing-and-being-seen}).

\subsection{Benefits of the Robot to the Homebound Student}
\label{sec:benefits}
All four participants reported that the benefits of the robot for them outweighed any drawbacks. After the end of the deployment, P1 said ``this robot is awesome'' and P4 said ``it was a great experience''. 
P1 said ``I wish I could use this robot again in the future'', and P3 said in Week~2 ``I want to keep using the robot''. The rest of this section details specific benefits of the robot to one or more of the homebound student participants. 


\subsubsection{Autonomy of Body Movement}
\label{sec:autonomy}
The robot's mobile base was especially important for P1 and P3, to whom it gave the ability to move around the classroom. 
This was only possible because the study team and school district worked with P1 and P3 to develop accessibility modifications, as described above. 

Before the deployment, both participants were typically pushed in wheelchairs while in the classroom, and both preferred driving the robot themselves: as P1 said in Week~2, ``I'm not just sitting there, letting other people do it for me''. 

One way that P3 used the new independence was to drive the robot out of the classroom in the middle of a class and down the school hallway to visit another teacher. 
Later, in Week~6, their mother described it as ``escaping the classroom'', and said that P3 was ``so happy'' about it. 

\subsubsection{Social Connection}
All four participants reported benefits of the robot related to connecting with people in the classroom. 
P1 in Week~2 said that interacting with people and meeting new people were the best parts of the week. 
For the first time in months, P2 saw their friend and their favorite teacher. 
In Week~1, P2 expressed that this was a good way to get to know people again so returning to school in person would not be so overwhelming. 
They brought their dog to the camera, facilitating connections with people in the classroom. 
P3 said in Week~1, ``the best thing is I got to see my friends''. 
Their mother said in Week~2 that the robot brought P3 extra attention from classmates, which P3 ``loves''. 

Before the deployment, P4 was ``very isolated'', according to one of their parents. 
Throughout the deployment, several of their friends helped to move the robot between classrooms, but spoke to P4 only briefly. 
In Week~1, however, P4 made a new friend who did the same type of drawing as P4, and was planning to go to the same high school for the arts. 
One of P4's parents was ``filling up with tears'' listening to their interaction, and said P4 ``[came] alive''. 
P4 believes the meeting might not have happened if they were attending in person: ``I'm not someone people notice when I'm at school ... [but] for the first time ever people actually talked to me''. 
In Week~13 they also noted a social benefit to their classmates and teachers: ``I was able to be seen by people in the class after being home for so long being sick.''

\subsubsection{Managing Anxiety}
Both P2 and P4 found that while they were previously anxious and distracted about what other students might be talking about, using the narrow field of view of the robot's camera helped them to focus on the front of the classroom. 
P2 in Week~1 used the robot interface's zooming function so that they could only see the class' TV screen during a presentation. 
P4 predicted a similar benefit during their training session, then confirmed that it came to pass during the deployment. 
P1, however, wanted a wider field of view for the camera. 

\subsubsection{Managing Fatigue}
Due to their medical context, P3 found it useful to log in or out of the robot at any time that they chose. 
In Weeks 2 and 4, their mother and brother said P3 typically only had about an hour of energy for classroom activities. 
Previously P3 rode a bus to school, ate breakfast in the cafeteria, and socialized, rendering them exhausted even before classes started. 
Instead of needing to commit to a six-hour day of school, P3 attended for one hour per day via the robot. 
Being able to log out was also a benefit when P3 had a medical emergency during class in Week~3. 
P2 was also able to leave class by logging out when they suddenly felt sick in Week~1. 

\subsubsection{Safety}
For P3 and P4, attending school via the robot was safer than attending in person. 
P3 had missed several months of school in the past after contracting a respiratory illness from others at school; treatment included a stay in the hospital's intensive care unit. 
In Week~6 they said they preferred attending via the robot rather than in person. 
P4 said that when they used to come to school with a cane, a student pretended to trip them by putting out their foot. 
After that, P4 had someone walk with them for safety. 
Participating via the robot protected these participants from illness and falls. 

\subsubsection{Advocacy}
During the deployment, P4 was placed in some advanced classes for the first time, by coincidence: the classrooms were near each other, and were chosen because it was easy to move the robot between them. 
P4's parents believed that the school tended to underestimate P4's fitness for advanced classes because they were in the school district's homebound student program, and that this placement led to their abilities being noticed. 

\subsection{Identified Challenges and Possible Solutions}
\label{sec:challenges-and-solutions}
Self-reports from study participants, their parents, and teachers revealed challenges with the deployment, and sometimes included suggested or improvised solutions. 

\subsubsection{Hearing and Being Heard}

Interview responses emphasized that classroom noise---i.e., other people talking---made it hard for classmates and teachers to hear the participant, and also for the participant to hear their classmates and teachers. 
People in the classroom had to repeat several times what they were saying to P3 in Week~2. 
Similarly, in Week~7 when P3 was telling everyone in class that their birthday had been over the weekend, nobody could understand anything besides the word ``Sunday''. 
Sometimes people coped with these difficulties via proximity: P3's teacher put their head up to the robot, and P4 drove up to each person they wanted to talk to, but the other person still needed to speak loudly. 
One of P2's teachers noted that ``we are relying so much on the speaker and microphone.'' 
P2's TA said that ``we have been able to hear [P2] just fine'' when the rest of the class is quiet and listening. 

P3 also found that listening to music was difficult when they attended a Christmas concert in Week~6. 
This was probably because the telepresence system classified the music as background noise and tried to filter it out. 
The concert took place outside the classroom, but this difficulty might also occur with classroom activities involving music. 

All four participants also experienced technical problems with the audio, especially at the beginning of their deployments. 
P1's audio was not reaching the classroom for all of Week~2; P1 described it as frustrating because they would try to answer a question, but others in the classroom thought P1 was ignoring them. 
P2 spent the first 20 minutes of their first day in remote class troubleshooting the same issue: the class could not hear them. 
P4 experienced an issue in the other direction: they could not hear the classroom for the first 30 minutes of their first day of attending class remotely. 
Other technical issues with the audio were more sporadic or partial: P3's teacher reported being able to hear P3 ``from across the room'' in Week~2, but then P3's audio ``faded in and out'' and lost connectivity while going across campus in Week~3. 

Sometimes workarounds were improvised when the robot's audio was not working. 
P2, their classmates, and one of their teachers held up handwritten signs to the camera. 
(One of the messages was not for troubleshooting; instead, it said: ``We miss you!'') 
When one of P3's classmates understood what P3 had said, they repeated it aloud for the rest of the class to hear. 
P3's mother texted the teacher directly, as did P4. 
P4 also called one of their teachers via the classroom phone. 

Solutions that could be integrated into the robot or its user interface were also suggested. 
P3's mother suggested a text-to-speech system, which could mitigate both P3's difficulty speaking clearly and issues with network latency and audio transmission quality. 
P4 and one of P2's teachers suggested that text could be displayed on the robot's screen, bypassing the audio channel. 

Another issue was the participants' uncertainty about how loud their voices were in the classroom. 
P2 in Week~1 tried to talk ``but everybody was talking and I don't know how loud I am.'' 
They were worried about being too loud if they ``yell[ed]'' and too quiet if they ``whisper[ed]''. 
These fears were realized in Week~2 when P2 was accidentally too loud when they were supposed to be quiet, drawing unwanted teacher attention to the group---the group was consequently upset with P2. 
P3 and P4 tested a prototype system for modeling and displaying feedback about speech intelligibility, as described by \cite{rueben2021long}.

\subsubsection{Seeing and Being Seen}
\label{sec:seeing-and-being-seen}
It was often difficult for the participants to view content being shown on a screen at the front of class. 
Glare was a problem for P2, and they had to use the interface's zoom feature when they were positioned at the back of the classroom. 
Another student's head sometimes blocked their view. 
P4 found that projected content was too bright to view, and that it was too dark to navigate with the downward-facing camera when the classroom lights were off during video projection. 
P4 suggested a ``camera shade'' to make the projector image less bright, and a light to illuminate around the robot's base while navigating. 

P4 and especially P2 struggled to get the teacher's attention by raising their hands. 
P2 felt frustrated in Week~1 when they raised their hand to volunteer for something, but was not seen even after leaning backward and forward and waving their hand. 
P4 felt unsure about how to raise their hand because they did not feel the teacher would notice; the robot was positioned at the front of the class, but to the side. 
P2 also found it difficult to tell when people were paying attention to them; sometimes P2 thought they were being talked to when in reality the teacher or other student was talking to someone behind the robot. 
This made P2 feel lonely. 

Sometimes when P2 raised their hand, a classmate noticed and alerted the teacher---``[P2] is raising [their] hand!''---and the teacher called on P2. 
P2 also once held up a piece of paper that said, ``hi!'' 
P2 suggested that the robot could physically raise something, as that is what teachers are looking for. 
One of P2's teachers also suggested a non-verbal signal, such as a light. 
P4 suggested that the robot could increase its height so that if it were in the back of the classroom, the teacher could see it. 

For most of their deployments, P1 and P4 took advantage of the interface button that turns off the camera observing their face. 
Prior to the deployment P4 said, ``it might take the first day or two to work up the courage to show my face'', and ended up doing so only with a small group of friends.

P1 and P4 both chose to display still images on the robot's screen as substitutes for their faces. 
P1 used a photo of a sports car they liked, and received compliments about it from classmates. 
P4 used two of their own drawings, which also drew compliments as well as questions from their classmates. 
When asked about personalizing the robot, P1 wanted to paint it blue and add some tattoos, to ``style it, make it into a fashionable car''. 
P4 suggested a feature for displaying emoticons on the robot's screen, either to express how they were feeling or to joke around. 

\subsubsection{Moving Around}
Navigating the robot was sometimes difficult for the participants, even after accessibility modifications were made to the user interface. 
P3 had to drive through narrow spaces, as their classroom had tight spacing and multiple teacher's aides walking around. 
P4 had a better experience in one of their classrooms: the teacher had set up red lines indicating where desks should go, and had told the other students to move their backpacks and feet to facilitate the robot. 
P3 also experienced network latency, which caused them to overshoot when turning to look at something in the classroom and when following someone down the hall. 
Finally, P2 found it disorienting when the robot was not in the expected position upon one login. 

One approach to addressing the navigation difficulties is to have someone in the classroom help the remote student. 
Sometimes a teacher provided such help, but we encouraged them to choose a student for this purpose, whom we called the ``bot buddy''. 
Bot buddies could serve as eyes and hands---e.g., for P4, they sometimes gave verbal navigation commands (``go a little to the right'') and at other times simply pushed or carried the robot. 

P4 also suggested increasing the robot's top speed to keep up with people walking down the halls, as well as adding a rear-facing camera for backing up. 

Participants' ability to attend activities outside the classroom was limited to places with a wireless network that had been set up on the robot. 
P2 missed a field trip to a courthouse for this reason. 
P3 attended Special Olympics, a cultural event in the school library, and went with the class to the school's community garden and baseball field, but experienced network interruptions and poor signal strength. 

\subsubsection{Classroom Management}
Participants were sometimes unable to participate fully in classroom activities. 
When the activities were accessible via a computer, such as when the students in the classroom were using Google Classroom via Chromebooks, this needed to be set up for the remote participant. 
P2 was participating fully in a history assignment in Week~1 because Google Classroom was set up for them, whereas P4 was not receiving credit for work they did in Week~1 because Google Classroom was not yet set up. 
There was also a communication gap between teachers and participants about what would happen in class. 
P2 often felt unsure about what they should be doing during their deployment. 
P3's mother said it would help to have ``more warning'' about the day's activities ``before just teleporting in''. 
For P4, information about in-class activities was posted on Google Classroom, but P4 was not given access for the first part of Week~1. 
Finally, participants were not able to participate in hands-on activities; for example, one of P4's teachers in Week~3 said, ``we were doing ... a `mock' assembly line and I could not find a way for [P4] to participate.'' 

It may have been partly due to the issues described above that P2 and P4 said that home school was more effective for learning than attending class via the robot. 
P2 felt bored with the robot, whereas with home school they knew what to do. 
P4 described home school as ``very fast paced'' and better for ``[learning] something fast to get caught up''. 

Another way to enable more participation by the remote student is by implementing a system for participation that includes every student. 
For example, one of P2's teachers drew popsicle sticks marked with seat numbers to select students for an activity; when P2's robot was positioned such that they had a seat number, they were called on, but on another day they did not have a seat number, and were not called on. 
P3's teacher had a policy that everybody in class got a turn for each activity, and P3 was included in and benefited from this policy. 

Teachers sometimes had difficulties maintaining discipline or control over remote participants. 
P1 often fell asleep during class; they were asleep for at least half of class on four out of the eight days they attended. 
Three factors made it possible for this to remain undetected by the teacher: P1 was not showing their camera feed, their snoring was too quiet for the microphone to pick up (when their audio was working), and they woke up and answered whenever someone asked them a direct question. 
One day in Week~3 after mostly sleeping through an hour of class, P1 said of their teacher, ``I don't think she knows I'm asleep'' and, ``she doesn't know what I'm doing.'' 

As already mentioned in Section~\ref{sec:autonomy}, P3 drove the robot out of the classroom without permission several times. 
P3's teacher also had another difficulty: P3 often logged in to the robot before class, and drove up to the teacher while the teacher was working. 
The teacher then had to go out into the hallway or even to the library to get away. 


\subsubsection{Negative Interactions with Classmates}
P2 and P4 both had negative interactions with some of their classmates. 
P2 had to leave class one day in Week~1 as a result of such interactions. 
Their bot buddy was absent, so another student whom P2 did not know well was the bot buddy. 
That student placed P2 at a table with ``the five most popular girls in class'', including the student who had previously posted negative things about P2 online. 
After a few ``really awkward'' minutes, P2 felt very sick and left the room; their mother came over to the laptop to get the teacher's attention and ended the robot session for the day. 

P4 experienced several types of negative interactions with classmates. 
In Week~1 a teacher put a school spirit shirt on the robot, but then a student took a photo of it without P4's permission. 
This discouraged P4 from showing their face on the robot's screen; they expressed worry that if they did, students would take pictures of it and post them on social media. 
In Weeks 1 and 2 students were waving their hands in front of the robot's camera and dancing in front of the robot. 
This did not feel like a greeting to P4; P4 felt unacknowledged, and found it ``distracting and overwhelming''. 
P4 also felt anxious that certain students would break the robot. 
When trying to drive down the hallway in Week~2, a student intentionally walked right in front of the robot. 
P4 reported that these behaviors had partly subsided by the end of their deployment. 


\section{Discussion}
This section highlights and discusses key findings about (1) homebound students' experiences, (2) recommendations for telepresence systems, (3) real-world deployments, and (4) future studies of mobile remote attendance in \mbox{K-12} classrooms; it also discusses (5) some limitations of our study.

\subsection{Multi-Dimensional Variance Among Deployments}
\label{sec:inter-student-variance}
Our four participants and the contexts of their deployments varied on multiple variables that constituted important factors in shaping their experiences. 
Participant variables that proved important included the participants' medical contexts, past experiences with in-person attendance, personalities (e.g., desire to socialize vs. to keep to oneself), home environment and network quality, and the personalities of parents and other caregivers. 
Deployment context variables that proved important included how effectively the teacher(s) integrated the remote student into the classroom and learning activities, how classmates treated the remote student, and network quality. 
The benefits and challenges that each participant experienced with the robot also varied, often in connection with personal and contextual differences. 

Furthermore, many of the differences among participant experiences were unaligned with one another. 
None of the following four segmentations of the participants are the same: P2 and P3 were outgoing and talkative, whereas P1 and P4 were reserved; P2 and P4 suffered from anxiety or fear from negative experiences at school in the past; P1 and P3 could not move their bodies around the classroom without the robot; and P1 and P4 preferred not to show their faces. 
This broad diversity even in a small sample suggests that students may not be able to be clustered into a small number of types; instead, key variables can vary independently, yielding many different combinations and meriting personalization toward significantly improved educational outcomes and user, teacher, and peer enjoyment of the process.

\subsection{Telepresence System Design}
This study documented benefits of mobile remote attendance for all four of the participants. 
The benefits can be clustered by whether they depended on mobility---i.e., on the homebound student's ability to drive the telepresence system around the classroom. 
Benefits of remote attendance that did not seem to depend on mobility included social connection and a suite of benefits for managing medical situations gracefully, flexibly, and effectively---in particular, managing anxiety and fatigue, handling emergencies, and staying safe from physical injury and contagious illnesses.
The main benefit of mobility seemed to be for students with limited movement autonomy (in our study this was due to their medical conditions) to be able to ``move'' on their own via the robot. 
This suggests that, especially for students without physical mobility difficulties, a simpler telepresence system that cannot be driven around the room might provide many benefits at reduced difficulty and cost. 
Options for such systems include a laptop computer with video conferencing software or a desktop robot with a camera and display that the homebound student can pan and tilt.
Either solution could be moved from place to place by teachers or classmates to facilitate inclusion and learning.
Future work should measure what else is lost when substituting tabletop systems for mobile ones, however; for example, mobility may have given our participants more control over choosing, initiating, and avoiding social interactions. 

We also documented various challenges with the telepresence robot system that resulted in worsened or negative experiences. 
Recurring difficulties across participants included hearing and being heard in noisy classrooms, reading content from a bright screen, getting the teacher's attention from far away, keeping up with others while driving the robot, and network latency with the audio and while driving the robot or adjusting the robot's camera angle.
In their study of 19 homebound students, \cite{ahumada2019going} documented many of the same difficulties and made recommendations for addressing each.
Also, \cite{jakonen2021mediated} studied ways for people in the classroom to make sure the robot operator can see a particular person or object. 
Further design work to address these challenges would improve future deployment experiences.

\subsection{Real-World Deployment Procedure}
In conducting this study we learned about steps that a school or school district would need to take in order to implement mobile remote attendance for homebound students, as discussed in Section~\ref{sec:deployment-procedure}. 
The first such step is obtaining agreement to start a mobile remote attendance program from all the stakeholders in the school system, including teachers and the teachers' union, technical support staff, principals and vice principals, and district-level administrators. 
During the setup of our study, for example, there were concerns about whether recordings from the robot's camera could be used by principals to evaluate teacher performance. 
The model of telepresence robot acceptance prepared by \cite{han2020use} is relevant in considering these issues. 

The second step regards procuring the telepresence technology. 
Telepresence robots could be owned by a school or shared within a school district. 
Additionally, we recommend our strategy of using a dedicated laptop for each telepresence session, also requiring resources and setup. 
Some telepresence robot companies have gone out of business; schools would need to make sure that any services that the robot depends on from the manufacturer would remain available for the full lifetime of the robot. 

The third step is actively monitoring the student body for students who might benefit from mobile remote attendance, and helping them and their families to decide whether to participate in such a program. 
We recommend a continuous monitoring process because transitions between school, home, and hospital can be sudden and unpredictable, and it can be difficult to judge when a student will be at home and feeling well enough to attend class remotely for enough weeks to justify the effort of a deployment. 
\cite{soares2017mobile} provide relevant experiences to inform this process.
Students being considered for mobile remote attendance should be briefed on what the experience will be like, including potential benefits, potential risks, and likely difficulties, and should be given a choice between mobile remote attendance and at least one alternative. 
Parents and other caregivers should understand and willingly accept the additional responsibilities they would have for supporting the deployment, and the importance of setting up a home environment that is conducive to mobile remote attendance. 
\cite{newhart2017my} provide suggestions for creating a ``safe space'' both at home and in the classroom. 

The fourth step is training the homebound student and their caregivers on using the telepresence system, and setting up any adaptations for accessibility.
For our study, accessibility was managed by teachers on special assignment (TOSAs) from the school district who specialize in educational technology, while the research team provided the training. 

The fifth step is defining and assigning duties, and establishing communication protocols, such as in the critical component structure by \cite{johannessen2022student}. 
All the duties for a deployment should be listed and defined, such as providing live technical support for the robot, communicating class announcements to the remote student if they were absent, checking in with them weekly to identify problems, escorting the robot between classrooms, and even making sure the robot is charged, powered on, and positioned correctly by the start of each class session.
Responsible parties (individuals or groups) need to agree to be responsible for each of the duties.
Next, communication protocols should be established.
The homebound student and their caretakers should know whom to contact for each of the different types of issues: technical issues, accessibility issues, logistical questions, and issues accessing assignments and other course materials. 
Teachers should have alternative ways to communicate with the homebound student when the robot is not working.

The sixth step is the inclusion and integration of the remotely present student into the classroom, including making assignments accessible, introducing them to the class, actively including them in activities, and responding to harassment.
\cite{charteris2022virtual} provide a list of questions for teachers and administrators for promoting inclusion for remote students. 
Teachers should receive training on mobile remote attendance, including stories of real problems and best practices to mitigate them, and should meet the homebound student and their care team prior to the deployment.

It might be unrealistic to expect teachers to manage every aspect of the homebound student's remote attendance in the classroom.
Additional support could be provided by specially trained TOSA from the school district, who could be present with the robot in the classroom, especially at the beginning of the deployment, and check in periodically thereafter.

Best practices still need to be developed for all of the above steps. 
As discussed in Section~\ref{sec:inter-student-variance}, this is a complex challenge because of the multi-dimensional variance among homebound students; each student's deployment may require individuated attention and planning.

\subsection{Insights for Future Study Design}
Interviewing participants and their families weekly yielded rich information about contexts, situations, and thought processes. 
Also, respondents could think back over the previous week and select the stories that were most memorable and impactful. 
We also recommend audiovisual recordings (with appropriate consent) of the homebound participant and of the classroom, and especially of events that participants highlight during interviews, because the audio and video capture important details that are not mentioned in interview responses. 
For data analysis, we recommend that future work continue to treat each deployment as a separate context via a case study methodology or similarly relevant methods, and that a methodology like thematic analysis could improve the process for identifying themes. 

We make two additional recommendations for future studies, specifically studies of the usability of a new design feature, or of the effect of mobile remote attendance on a particular social or educational variable. 
First, such studies should be designed with the assumption that the following variables may have large effects on outcomes: how talkative, confident, or anxious the participant is, how well they know their teacher and classmates, what their existing positive or negative relationship is with a teacher or classmate, how often they will be attending class, what their accessibility needs are to drive the robot, and whether they have difficulties speaking or hearing. 
Eventually, to scale up study results, study participant recruitment should be restricted to subsets of this complex space, but given the variance in student needs, this will likely require a geographically broadly distributed data collection. Obtaining a representative sample across all relevant variables will remain a major challenge because of the heterogeneity of the user population. 

Second, studies should consider using a withdrawal design \citep{byiers_single-subject_2012}, i.e., the ABA (baseline-intervention-baseline) approach. Understanding the effects of transitions---e.g., between home instruction and remote attendance---will likely be relevant for many students due to hospital stays and other changes in context. 

\subsection{Limitations}
Recruiting participants for our study was incredibly challenging because it involved numerous components and agreements.  
We were fortunate to work with the students, families, teachers and schools who collaborated with our research team, but we were not able to scale up our study with the available resources.
Our study participants were different from one another along multiple dimensions that critically affected their experiences with mobile remote attendance---e.g., medical context, personality, and past experiences in the classroom.
With only four participants and such high variance, our dataset supports very limited generalization. 
Because of the realities of recruitment, our study design had only one condition; we did not directly compare mobile remote attendance to an alternative solution such as home school. 

The longest deployment was less than a semester long, so we were not able to study how attending remotely for longer time scales may affect a student's educational and social experience. 
Furthermore, we did not continue to interview participants after they returned to home school in order to assess any lasting effects.

Interview responses were sometimes provided by caretakers and not by the participants.  Specifically, P3's mother and brother provided responses because it was hard for P3 to speak at length, while P4's mother and father provided responses because they enjoyed elaborating on P4's experiences. We did not formally interview any of the teachers, classmates, or other stakeholders. 

Due to logistical or technical reasons, some audiovisual recordings were missing, as described in Section~\ref{sec:recordings-results-overview}. 
Consequently, 
we were not able to review some events mentioned in the interviews because the audiovisual recordings of them were missing.  

\section{Conclusion}
We deployed a telepresence robot and user interface with four homebound \mbox{K-12} students in their homes and classrooms, allowing them to remotely attend class while embodied in a telepresence robot. 
Their experiences and reflections were collected via interviews and audiovisual recordings. 
Our combined findings inform the field about benefits of, challenges with, and recommendations for future deployments of mobile remote attendance systems.

The four participants were different from one another along multiple dimensions so that no clusterings were apparent. 
These differences were important predictors of our participants' experiences with mobile remote attendance, suggesting that mobile remote attendance should be personalized to each student's unique situation. 

All four participants benefited from increased social interaction. 
Other benefits were only enjoyed by participants with certain needs: managing their anxiety or fatigue, staying safe from physical injury or illness, or experiencing autonomy of body movement in the classroom. 
Some of these benefits do not require a system that can be driven around the classroom; for others, such as social connection, the importance of mobility should be studied further. 

Challenges with the deployments led to two types of recommendations: limitations in hearing, seeing, and moving the robot around the classroom warrant improvements to the design of telepresence systems, while other challenges suggest priorities for managing deployments, such as ensuring the remote student is included in activities, accountable to the teacher, and treated with respect by classmates. 

We also contribute recommendations for a real-world deployment procedure. 
This study highlighted that mobile remote attendance takes additional time, effort, and coordination between stakeholders compared to traditional classrooms. 
Purchasing the technology is far from the only step---mobile remote attendance makes demands on the family, the home environment, classmates, teachers, and other school employees in ways that in-person instruction does not.
Researchers and schools need to work together to develop technologies and procedures that provide higher quality education to homebound students in a way that is sustainable for school systems and families.  

\backmatter


\bmhead{Acknowledgments}
The authors thank the school district administrators, teachers, TAs, participants, families, and classmates, all of whom made this complex field research possible. We also thank OhmniLabs for technical support of the robots. We thank Jessica Lupanow for helping with project management. 

\backmatter

\bmhead{Declarations}

\bmhead{Funding}
This work was supported by a National Science Foundation (NSF) National Robotics Initiative grant for ``Socially Aware, Expressive, and Personalized Mobile Remote Presence: Co-Robots as Gateways to Access to \mbox{K-12} In-School Education'', NSF IIS-1528121. H. Chen was supported by the University of Southern California (USC) Undergraduate Research Associates Program. H. Lee was supported by the USC Viterbi Summer Undergraduate Research Experience program. The Ohmni Robot platform and associated technical support were provided by OhmniLabs.

\bmhead{Competing Interests}
The authors have no competing interests to declare that are relevant to the content of this article.

\bmhead{Ethics Approval}
This study was approved by the University of Southern California Institutional Review Board (IRB), study number UP-18-00129.

\bmhead{Data Availability}
Due to their sensitive nature, neither the interview transcripts nor the audiovisual recordings may be shared outside of the research team.

%
%
%
%

\begin{appendices}

\section{Pool of Interview Questions}
\label{append:interview-question-pool}

\subsection{General Impressions}
\label{append:general-impressions}
\begin{itemize}
    \item Tell me a little about how it went this week in class. 
    \item How did you feel about your week?
    \item How did class go today?
    \begin{itemize}
        \item Tell me the story – what happened in class? Walk me through it.
        \item (Follow up by asking how they were involved in each part.)
    \end{itemize}
    \item Did you feel a bit shy on the first day of the week?
    \item Did you feel a bit differently as time went on?
    \item What were some of your favorite parts of being in class?
    \item Were there any times that were kind of awkward or uncomfortable? Describe.
    \item In what ways was it different than you expected?
    \item Is there anything you're worried or concerned about that you want to share?
    \item Anything else that might go wrong that you're worried about?
\end{itemize}

\subsection{Interactions with Classmates}
\label{append:interactions}
\begin{itemize}
    \item How did it feel interacting with the other people in your class?
    \item Were there times when you wanted to get someone's attention? What ways did you use to get their attention?
    \item Did you ever feel like people were missing your signals when you wanted their attention? Can you tell me an example or two?
    \item Did you think of ways to get students' attention? If so, what did you do?
    \item Did you ever feel like people gave you extra attention compared to what people usually give you (when you're not driving a robot)? Tell me a bit about that.
    \item Were there times when you didn't want to attract attention? What did you do to accomplish that? How did you stay in the background?
    \begin{itemize}
        \item Did people ever talk to you or put you on the spot when you intended to stay quiet and inconspicuous? Tell me a bit about that.
    \end{itemize}
\end{itemize}

\subsection{Relationships with Classmates}
\label{append:relationships}
\begin{itemize}
    \item How do you think it is going with your new classmates?
    \item Have you gotten to talk with anybody in class yet? Tell me about one of them:
    \begin{itemize}
        \item Do you remember their name?
        \item What did you talk about together?
    \end{itemize}
    \item How about some other students in the class?
    \begin{itemize}
        \item Of the people you have met in class so far, who would you most like to get to know more? Why?
    \end{itemize}
\end{itemize}

\subsection{Learning Experience}
\label{append:learning}
\begin{itemize}
    \item How easy or hard was it to pay attention in class compared to homeschool? Tell me more.
    \item Did you do any of the class activities? List all of them.
    \begin{itemize}
        \item (For each one...) How easy was it to participate in that? Why?
        \item Any other activities you did?
    \end{itemize}
\end{itemize}

\subsection{The Robot's Technical Abilities}
\label{append:technical}
\begin{itemize}
    \item Tell me a bit about how you think the robot worked in the classroom.
    \item How well could you \emph{hear} through the robot? What's an example of when you could hear very well? What about an example when it was hard to hear? Any more?
    \begin{itemize}
        \item How well can you hear at within 5ft? What's the farthest away you could hear from, do you think?
    \end{itemize}
    \item How well could you \emph{see} through the robot? Were there times when you couldn't see something (beside or behind or above or below) that you wanted to see? List as many as you can remember.
    \begin{itemize}
        \item What's your impression of how wide the robot's view is?
    \end{itemize}
    \item Did you ever feel like your voice volume was too loud or quiet for people in the classroom?
    \begin{itemize}
        \item (If so...) When was that? Tell me about it.
    \end{itemize}
    \item How easy or difficult was it to use the robot interface while you were in class?
    \begin{itemize}
        \item Tell me more about that.
    \end{itemize}
    \item Was the interface ever distracting or overwhelming?
    \begin{itemize}
        \item (If so...) Can you think of a change to the interface that would make it less overwhelming?
    \end{itemize}
\end{itemize}

\subsection{Ideas for Improving the Robot or Interface}
\label{append:improvements}
\begin{itemize}
    \item What extra features would you like the robot to have? Let's brainstorm some things you'd like to be able to do in the classroom. It could be fun stuff, stuff to help you do classroom activities, ways to chat better with your classmates, ...
    \begin{itemize}
        \item What would you put up on your screen?
        \item What would you do if you had an arm?
        \item What emotions would you want to express? 
    \end{itemize}
\end{itemize}

\subsection{Goals and Regrets}
\label{append:goals-and-regrets}
(During the deployment...)
\begin{itemize}
    \item What goals do you have for next week?
    \item How about \emph{big} goals for the end of the study?
\end{itemize}
(After the deployment...)
\begin{itemize}
    \item If you could attend class via the robot for another week, what would you make sure to do?
    \item Do you have any regrets from these four weeks? Anything you wish you had done differently?
\end{itemize}





\end{appendices}


\bibliography{main}

\end{document}